\documentclass[12pt,twoside]{article}
\usepackage{fleqn,espcrc1}
\usepackage{graphicx}
\usepackage{epsfig}
\usepackage{floatfig}

\newcommand{\AmS}{{\protect\the\textfont2
  A\kern-.1667em\lower.5ex\hbox{M}\kern-.125emS}}
\newcommand{\sqrts}{\mbox{$\sqrt{s}$}}
\newcommand{\sqrtsNN}{\mbox{$\sqrt{s_{_{\mathrm{NN}}}}$}}
\newcommand{\AuAu}{\mbox{$\mathrm{Au}+\mathrm{Au}$}}
\newcommand{\PbPb}{\mbox{$\mathrm{Pb}+\mathrm{Pb}$}}
\newcommand{\ppbar}{\mbox{$p\bar{p}$}}

\newcommand{\hminus}{\mbox{$h^-$}}
\newcommand{\Nhminus}{\mbox{$N_{h^-}$}}

\newcommand{\piminus}{\mbox{$\pi^-$}}
\newcommand{\kminus}{\mbox{$K^-$}}
\newcommand{\pbar}{\mbox{$\bar{p}$}}
\newcommand{\meanpt}{\mbox{$\langle p_\perp \rangle$}}
\newcommand{\pt}{\mbox{$p_\perp$}}

\newcommand{\gevc}{\mbox{${\mathrm{GeV/}}c$}}
\newcommand{\mevc}{\mbox{${\mathrm{MeV/}}c$}}
\newcommand{\TAA}{\mbox{$\mathrm{T}_{\mathrm{AA}}$}}
\newcommand{\siginel}{\mbox{$\sigma_{\mathrm{inel}}$}}
\newcommand{\nbincoll}{\mbox{$N_{\mathrm{BC}}$}}
\newcommand{\npart}{\mbox{$N_{\mathrm{WN}}$}}

\begin{document}
\initfloatingfigs

% declarations for front matter

%\begin{frontmatter}
\title{Negatively Charged Hadron Spectra \\ in Au+Au Collisions at $\sqrt{s_{NN}}$ = 130 GeV}

\author{M. Calder\'{o}n de la Barca S\'{a}nchez 
\address{Department of Physics, Yale University \\ P.O. Box 208120, New Haven CT 06520 }
for the STAR Collaboration.
\thanks{For complete author list see J. W. Harris, these proceedings.}}

% typeset front matter
\maketitle

\begin{abstract}
    Negatively charged hadron (\hminus) production in Au+Au collisions
    at BNL-RHIC is studied with the STAR experiment.  Results are
    presented on \hminus\ multiplicity, pseudorapidity and transverse
    momentum distributions at $\sqrt{s_{NN}}$ = 130 GeV.
\end{abstract}
%\end{frontmatter}

\section{Introduction}
Global observables such as the multiplicity and the inclusive single
particle transverse momentum (\pt) and pseudorapidity ($\eta$)
distributions of hadrons have been valuable tools in studying heavy
ion collisions. They represent the system at kinetic freeze-out, late
in the evolution of the system when hadrons no longer interact with
each other, and their momentum spectra do not change further.  These
final-state observables supply essential constraints on the possible
evolutionary paths of the system that can help establish conditions in
the early, hot and dense phase of the collision.  A discussion on
global observables
can be found in \cite{eskola}.
We summarize here results on the minimum-bias multiplicity, \pt\ and
$\eta$ spectra of negatively
charged hadrons.% (\hminus).

\section{Experiment and Analysis}
The STAR experimental setup for the first RHIC run is described in
\cite{star}.  The analysis is based on charged particle tracking in
the STAR Time Projection Chamber (TPC).  Triggering was achieved using
two hadronic calorimeters (ZDCs) in the very forward region and an
array of scintillator slats arranged in a barrel (CTB)
around the TPC.

Particle production was studied through the yield of primary negative
hadrons (\hminus), \textit{i.e.} \piminus, \kminus\ and \pbar\ 
including the products of strong and electromagnetic decays.  Low
momentum ($< 1 \gevc$) particle identification was done via $dE/dx$ in
the gas of the TPC.  Only negatively charged hadrons were studied in
order to exclude effects due to participant nucleons.  Charged
particle tracks reconstructed in the TPC were accepted if they
fulfilled requirements on number of points on the track and on
pointing accuracy to the event vertex.  The measured raw distributions
were corrected for acceptance, reconstruction efficiency,
contamination due to interactions in material, misidentified
non-hadrons, the products of weak decays, and track splitting and
merging.  The tracking efficiency was found by embedding simulated
tracks into real events at the raw data level, reconstructing the full
events, and comparing the simulated input to the reconstructed output.

For this analysis, the acceptance for tracks within the fiducial
volume having \pt\ greater than 0.3 \gevc\ is found to be 95\%.  The
tracking efficiency is found to be between 70--95\%, depending on the
particle \pt\ and the total multiplicity of the event.

%-------------------------------------------------------------------------------
%  Results and Discussion: dsigma/dNh-
%-------------------------------------------------------------------------------
\section{Results and Discussion}
\begin{floatingfigure}{75mm}
    \mbox{\epsfig{file=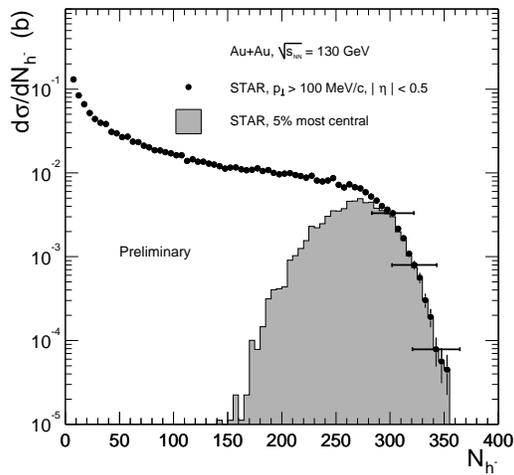, width=0.45\textwidth}}
       \caption{Multiplicity distribution
           of \hminus, shaded area is for the 5\% most central
           collisions.}
       \label{fig:hminus}
\end{floatingfigure}

Figure \ref{fig:hminus} shows the \hminus\ multiplicity distribution
for minimum-bias Au+Au collisions with $|\eta|< 0.5$ and $\pt > 100\ 
\mevc$. The data were normalized assuming a total hadronic inelastic
cross section of 7.2 barn for Au+Au collisions at \sqrtsNN\,= 130 GeV,
derived from Glauber model calculations \cite{glauber,dima}.

The systematic error on the vertical scale is estimated to be 10\% and
is dominated by uncertainties of the total hadronic cross-section and
the shape at low $N_{h^-}$.  The systematic error of 6\% on the
horizontal scale is depicted by horizontal error bars on a few data
points.

The shape of the distribution is dominated by the collision geometry:
large cross section at low multiplicity, corresponding to large impact
parameter, followed by a region of slowly falling cross section over a
wide range of multiplicity as the nuclei overlap, and a rapid decrease
for near head-on collisions, where the shape is determined by
fluctuations in the collision geometry and particle production process
given the finite detector acceptance. The distribution for the 5\%
most central collisions (360 mbarn), defined via ZDC coincidence, is
shown as the shaded area in Fig.~\ref{fig:hminus}.

%-------------------------------------------------------------------------------
%  Results and Discussion: 1/pt dN/dpt
%-------------------------------------------------------------------------------
Figure \ref{fig:pt}, upper panel, shows the \hminus\ \pt\ distribution
for the 5\% most central collisions at midrapidity ($|\eta| < 0.1$).
Only statistical errors are shown; the systematic errors are estimated
to be below 6\%.  The data are fit by a power-law function of the form
$(1/\pt)\, d\Nhminus/d\pt = A\, (1+\pt/p_0)^{-n}$.
As reference data sets, we also show the \pt-distribution of
negatively charged hadrons for central Pb+Pb collisions at \sqrtsNN\,=
17 GeV from NA49 \cite{na49} and for minimum-bias \ppbar\ collisions
at \sqrts\,= 200 GeV from UA1 \cite{ua1}, fitted with the same
function.
The UA1 invariant cross-section $E d^3 \sigma/d^3p$ reported in
Ref.~\cite{ua1} was scaled by $2 \pi / \siginel$, where $\siginel =
42$ mb.
The \pt\ spectrum measured by STAR reflects a systematic increase in
\meanpt\ compared to that from both central A+A collisions at much
lower energy (NA49) and \ppbar\ collisions at a comparable energy
(UA1).
\begin{floatingfigure}{80mm}
    \mbox{\epsfig{file=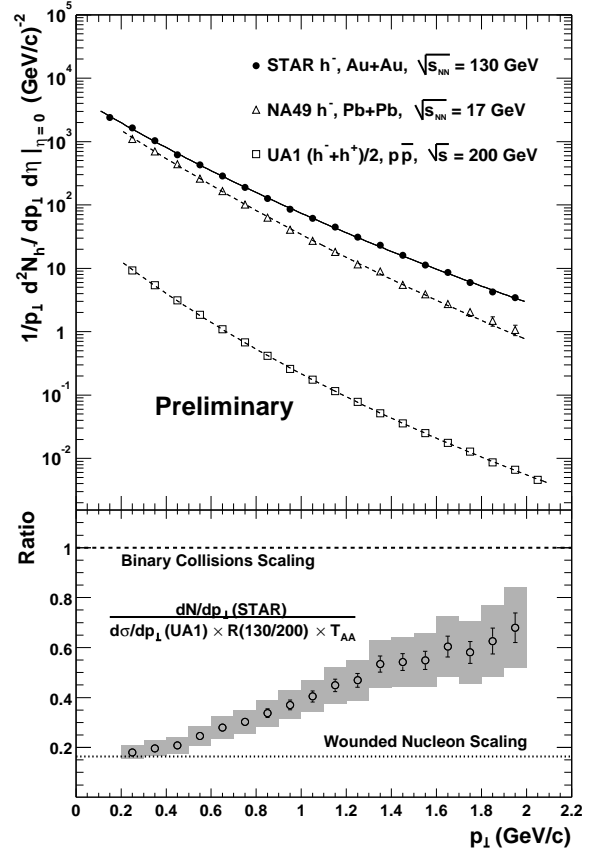, width=0.52\textwidth}}
        \caption{Upper panel: \hminus\ \pt-spectra for the 5\% most
            central collisions. NA49 central and UA1 data are also
            shown.  Lower panel: ratio of STAR and scaled UA1
            \pt-distributions.}
        \label{fig:pt}
\end{floatingfigure}

The lower panel of Fig.~\ref{fig:pt} shows the ratio of the STAR and
UA1 \pt-distributions. Since the UA1 distribution is measured at
\sqrts\,= 200 GeV, the invariant cross-section in each \pt-bin is
scaled by two factors for a quantitative comparison to the STAR data:
\textit{(i)} $R(130/200)$, the \pt-dependent ratio of the \hminus\ 
yields in \ppbar\ collisions at \sqrts\,= 130 and 200 GeV, and
\textit{(ii) } \TAA = 26 mb$^{-1}$, the nuclear overlap integral
\cite{taa} for the 5\% most central Au+Au collisions.  See
\cite{jamie} for further details.

There are two simple predictions for the scaled ratio. The study of
lower energy collisions has shown that the total pion yield due to
soft (low \pt) processes scales as the number of participants, i.e.~,
``wounded'' nucleons (\npart) in the collision, (e.g.~\cite{na49}).
The scaled ratio in this case is 0.164 assuming 172 participant pairs
\cite{dima} and a mean number of binary collisions (\nbincoll) of
$\nbincoll\ = \siginel\, \TAA = 1050$ for the 5\% most central Au+Au
events.  We assume $\siginel\ = 40.5$ mb for \ppbar\ collisions at
$\sqrts = 130$ GeV. If hadron production is due to hard (high \pt)
processes and there are no nuclear-specific effects (see below), the
hadron yield will scale as \nbincoll, which is proportional to the
nuclear overlap integral \TAA. In this case the value of the ratio is
unity.  There are important nuclear effects which should alter the
scaling from these simple predictions, including \textit{e.g.} initial
state multiple scattering \cite{cronin}, jet quenching \cite{jetq},
and radial flow \cite{rflow}. Each of these exhibits characteristic
features as a function of \pt\ and system size.  The scaled ratio
shows a strong \pt\ dependence, starting close to \npart\ %Wounded Nucleon
scaling at low \pt\ and approaching \nbincoll\ %Binary Collision
scaling, not reaching the latter in the measured \pt\ range even
including the errors from the $R(130/200)$ and \TAA\ scaling shown in
gray.  This behaviour is consistent with the presence of radial flow,
as well as the onset of hard scattering contributions and initial
state multiple scattering with rising \pt.  The shape of the
distribution at higher \pt\ is discussed in \cite{jamie}.

The \hminus\ density at midrapidity for $\pt > 100\, \mevc$ and
$|\eta| < 0.1$ is $dN/d\eta|_{\eta = 0} = 253 \pm 1_{stat.} \pm
16_{syst.}$.  Extrapolation to $\pt$ = 0 yields $dN/d\eta|_{\eta = 0}
= 275 \pm 1_{stat.} \pm 18_{syst.}$.  Assuming an average of 172
participant pairs per central Au+Au collision, this corresponds to
$1.60 \pm 0.13$\,\hminus\ per participant nucleon pair per unit of
pseudorapidity. This is a 35\% increase over \ppbar\ collisions at the
same energy \cite{ua5}.  Comparison with central \PbPb\ collisions at
SPS \cite{na49} shows an increase in the \hminus\ yield per
participant of $\sim$ 49\% at RHIC.

%-------------------------------------------------------------------------------
%  Results and Discussion: dN/deta
%-------------------------------------------------------------------------------

\begin{figure}[htb]
    \vspace{-1\baselineskip}
\begin{minipage}[t]{75mm}
    \includegraphics[width=1.15\textwidth]{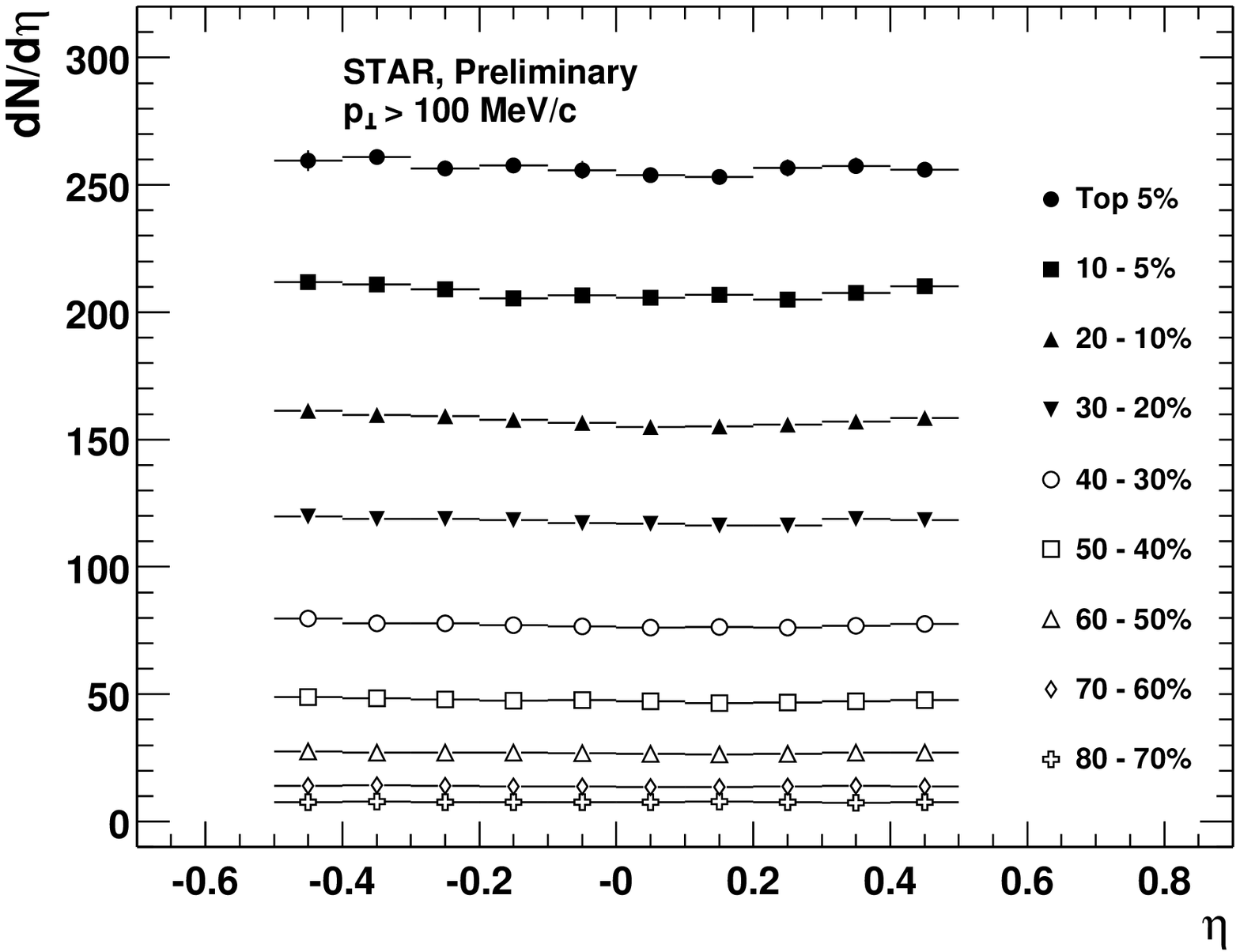}
        \caption{\hminus\ $\eta$ distribution for different centralities.}
        \label{fig:etacent}
\end{minipage}
\hspace{\fill}
\begin{minipage}[t]{75mm}
    \includegraphics[width=.9\textwidth]{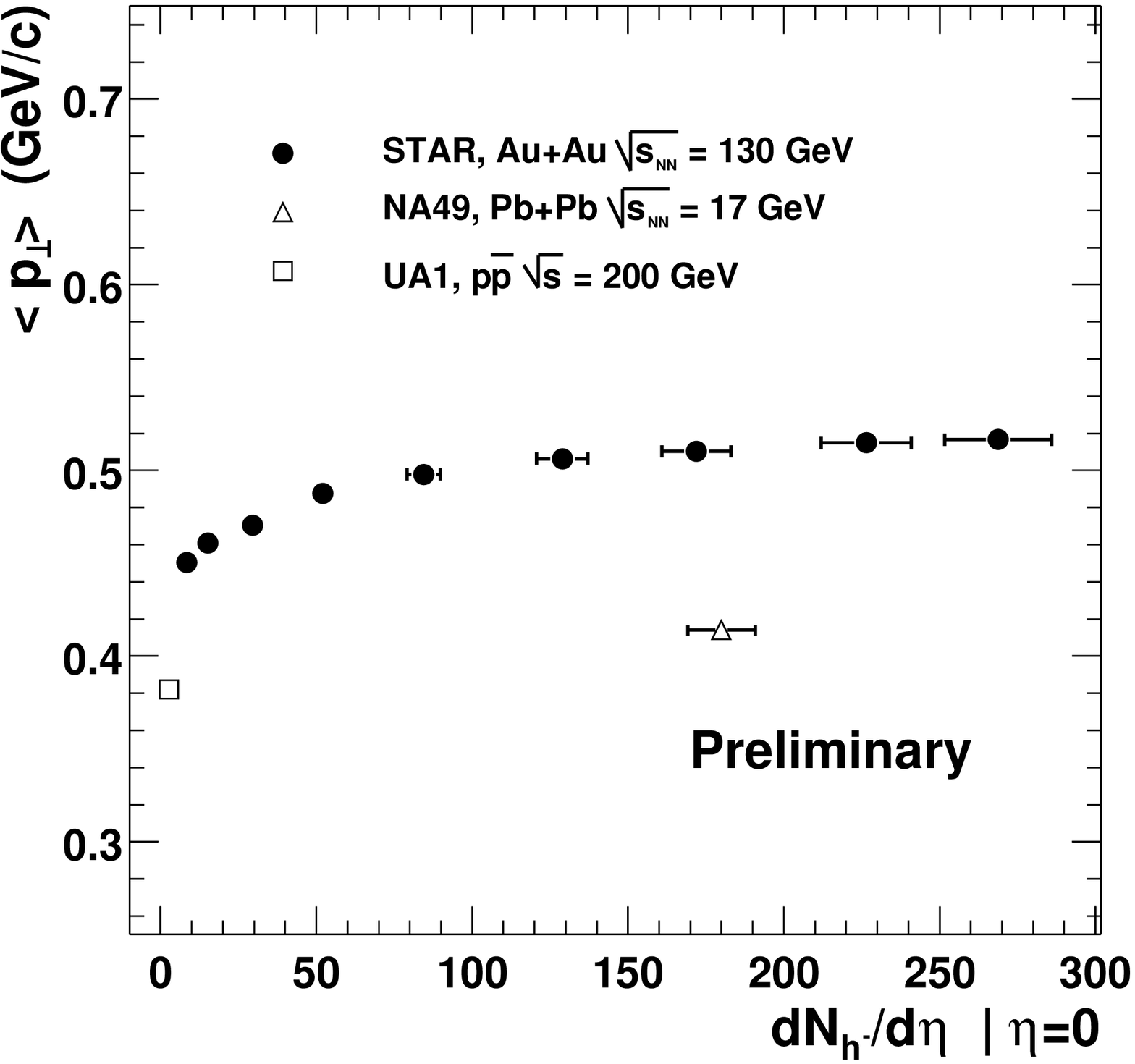}
        \caption{Centrality dependence of \meanpt.}
        \label{fig:meanptcent}
\end{minipage}
\end{figure}

Fig.~\ref{fig:etacent} shows the centrality dependence of the $\eta$
distribution.  One sees the expected rise in particle yield with
collision centrality.  Preliminary analysis shows very little
difference in the shape of the distribution at midrapidity with
increasing centrality.
Using again a power-law fit to the \pt\ spectra we obtain \meanpt\ 
which is shown in Fig.~\ref{fig:meanptcent} as a function of
centrality.  Errors on the horizontal scale are systematic.
We see an increase in \meanpt\ of 15\% from the most peripheral to the
most central events.  The values for the UA1 and NA49 reference data
sets are also shown.

%---------------------
\vspace{-1\baselineskip}
\section{Conclusion}

We have presented preliminary results on charged particle production
at RHIC.  We find $dN_{h^-}/d\eta$ per participant in central \AuAu\ 
collisions at \sqrtsNN\,=130 GeV increases by 35\% relative to \ppbar\ 
and 49\% compared to nuclear collisions at \sqrtsNN\,= 17 GeV. The
\pt\ distribution is harder than that of the reference systems for the
\pt\ region up to 2 \gevc, with the most central events having the
highest \meanpt.  Scaling of produced particle yield with number of
participants shows a strong dependence on \pt, with Wounded Nucleon
scaling achieved only at the lowest measured \pt.


\begin{thebibliography}{99}
\bibitem{eskola} K.~J.~Eskola, these Proceedings.  See also J.-P.~Blaizot, Nucl.~Phys.~A \textbf{661}, 3c (1999).
\bibitem{star}  F. Reti\`{e}re, these Proceedings.  K.H.~Ackermann \textit{et al.}, Nucl.~Phys.~A \textbf{661}, 681c (1999).
%\bibitem{hijing} X.N.~Wang and M.~Gyulassy, Comput.~Phys.~Commun.~\textbf{83}, 307 (1994). 
\bibitem{glauber} A.J.~Baltz, C.~Chasman, and S.N.~White, Nucl.~Instr.~Meth.~A \textbf{417}, 1 (1998).
\bibitem{dima} D.~Kharzeev and M.~Nardi, Phys.~Lett.~B \textbf{507} 121 (2001) nucl-th/0012025.
\bibitem{na49} H.~Appelsh\"auser \textit{et al.}, Phys.~Rev.~Lett.~\textbf{82}, 2471 (1999).
\bibitem{ua1} C.~Albajar \textit{et al.}, Nucl. Phys.~B \textbf{355}, 261 (1990).
%\bibitem{ua5x} G.J.~Alner \textit{et al.}, Z.~Phys.~C \textbf{32}, 153 (1986). 
\bibitem{jamie} J.C.~Dunlop, these proceedings.
\bibitem{taa} K.~J.~Eskola, K.~Kajantie, and J.~Lindfors, Nucl.~Phys.~B \textbf{323}, 37 (1989).
\bibitem{cronin} D.~Antreasyan, Phys.~Rev.~D \textbf{19}, 764 (1979).
%\bibitem{shadowing} V. Emel'yanov \textit{et al.}, Phys.~Rev.~C \textbf{61}, 044904 (2000).
\bibitem{jetq}  X.N.~Wang, Phys.~Rev.~C \textbf{58}, 2321 (1998).
\bibitem{rflow} P.~F.~Kolb, J.~Sollfrank, and U.~Heinz, Phys.~Rev.~C \textbf{62}, 054909 (2000).
\bibitem{ua5} G.J.~Alner \textit{et al.}, Z.~Phys.~C \textbf{33}, 1 (1986). 
%\bibitem{phobos} B.B.~Back \textit{et al.}, Phys.~Rev.~Lett.~\textbf{85}, 3100 (2000).
\end{thebibliography}
\end{document}